\documentclass[%
 reprint,
superscriptaddress,
 amsmath,amssymb,
 aps,
prb,
longbibliography,
]{revtex4-2}

\usepackage{graphicx}%
\usepackage{dcolumn}%
\usepackage{bm}%
\usepackage{siunitx}
\usepackage[version=4]{mhchem}
\usepackage[hidelinks]{hyperref}%
\usepackage{cleveref}%

\usepackage[american]{babel}
\begin{document}
\preprint{APS/123-QED}

\title{Coupling-dependent Antiferromagnetic-Ferromagnetic Ordering in Pinwheel Artificial Spin Ice}

\author{Anders Str\o mberg}
 \email{anders.stromberg@ntnu.no}
 \affiliation{Department of Electronic Systems, Norwegian University of Science and Technology, Trondheim, Norway.}

 \author{Einar Digernes}
  \affiliation{Department of Electronic Systems, Norwegian University of Science and Technology, Trondheim, Norway.}

\author{Rajesh Vilas Chopdekar}
 \affiliation{Advanced Light Source, Lawrence Berkeley National Laboratory, Berkeley 94720, California, USA}

\author{Jostein Grepstad}%
 \affiliation{Department of Electronic Systems, Norwegian University of Science and Technology, Trondheim, Norway.}%
\author{Erik Folven}%
 \affiliation{Department of Electronic Systems, Norwegian University of Science and Technology, Trondheim, Norway.}%

\date{\today}

\begin{abstract}
Nanopatterned magnetic thin films offer a platform for exploration of tailored magnetic properties such as emergent long-range order.
A prominent example is artificial spin ice (ASI), where an arrangement of nanoscale magnetic elements, acting as macrospins, interact via their dipolar fields.
In this study, we discuss the transition from antiferromagnetic (AF) to ferromagnetic (FM) long-range order in a square lattice ASI as the magnetic elements are gradually rotated through \SI{45}{\degree} to a ``pinwheel'' configuration. 
The AF--FM transition is observed experimentally using synchrotron radiation x-ray spectromicroscopy and occurs for a certain rotation angle of the nanomagnets, dependent on the dipolar coupling strength determined by the separation of the magnets in the lattice.
Large-scale magnetic dipole simulations show that the point-dipole approximation fails to capture the correct AF--FM transition angle.
However, excellent agreement with experimental data is obtained using a dumbbell-dipole model, which better reflects the actual dipolar fields of the magnets.
This model also explains the coupling-dependence of the transition angle, another feature not captured by the point-dipole model. 
Our findings resolve a discrepancy between measurement and theory in previous work on ``pinwheel'' ASIs and establish the coupling-dependence of the AF--FM transition. %

The revised dipole model, with a more accurate representation of the stray field, offers more precise control of magnetic order in artificial spin systems.

 \end{abstract}

                              
\maketitle

Artificial spin ice (ASI) systems\citep{Wang2006} composed of single-domain nanomagnets, acting as macrospins, exhibit a wide range of exotic behavior\citep{Mengotti2011,Skjaervo2020,May2021,Digernes2021}. 
Recent work suggests that these systems are key to emergent device technologies such as physical reservoir computing\citep{Jensen2018,Gartside2022}. 
The interaction between the macrospins in an ASI is mediated by the stray fields of the individual magnets, creating a complex energy landscape sensitive to the exact shape of the elements and the geometric arrangement of the ensemble\citep{Skjaervo2020,Digernes2020}. 
With precise nanofabrication, this sensitivity enables effective tuning of magnetic order.
Accurate modeling of the magnetic interactions in these dipolar-coupled metamaterials is essential to the analysis of emergent behavior, such as long-range magnetic order.

Previous research has established the predominance of long-range ferromagnetic (FM) order in a \SI{45}{\degree} ``pinwheel'' ASI\citep{Macedo2018}, in contrast to the antiferromagnetically (AF) ordered  ground state of the square ASI\citep{Wang2006}.
The FM-ordered pinwheel ASI is a meta\-material with several exotic emergent properties such as dynamic chirality\citep{Gliga2017b}, unique domain-wall topologies\citep{Li2019}, and Heisenberg pseudo-exchange\citep{Paterson2019}.

\begin{figure}
    \centering
    \includegraphics[width=\columnwidth]{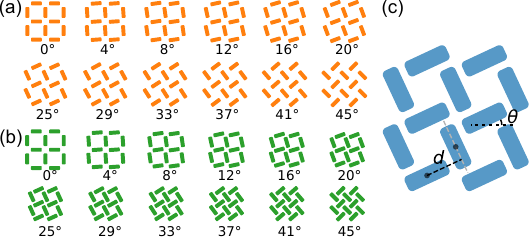}
    \caption{
    Geometric configuration of square-to-pinwheel structures. 
    (a) Rotation of elements from the square $\theta =\SI{0}{\degree}$ ASI (top left) to the typical $\theta =\SI{45}{\degree}$ pinwheel ASI (bottom right). 
    (b) The same series as in (a), but with the lattice compressed to compensate for decoupling. 
    (c) Schematic showing the geometric parameters of the systems. 
    }
    \label{fig:geometry}
\end{figure}

In this Letter, we investigate the AF to FM ordering transition we observe when going from a square to a pinwheel ASI\citep{Macauley2020,Massouras2020}.
Specifically, we examine modifications of the square ASI by gradually rotating each nanomagnet on the square lattice from $\theta = \SI{0}{\degree}$ to $\theta = \SI{45}{\degree}$, see \cref{fig:geometry}.
We observe the long-range magnetic order of these meta\-materials using x-ray magnetic circular dichroism (XMCD) spectromicroscopy at the Advanced Light Source (ALS). 
A predominance of FM order is seen for the ASIs with a $\theta$ close to the \SI{45}{\degree} pinwheel configuration.
We find a transition angle for the AF--FM ordering dependent on the separation of the nanomagnetic elements, at odds with previous reports relying on a simple point-dipole model\citep{Macedo2018,Macauley2020,Massouras2020}.
For a more accurate representation of the stray fields, we invoke a dumbbell-dipole model, which is found to accurately capture the  dependence of the AF--FM transition angle on the nanomagnet separation\citep{Castelnovo2008}.

\begin{figure*}[h!t]
    \centering
    \includegraphics[width=\textwidth]{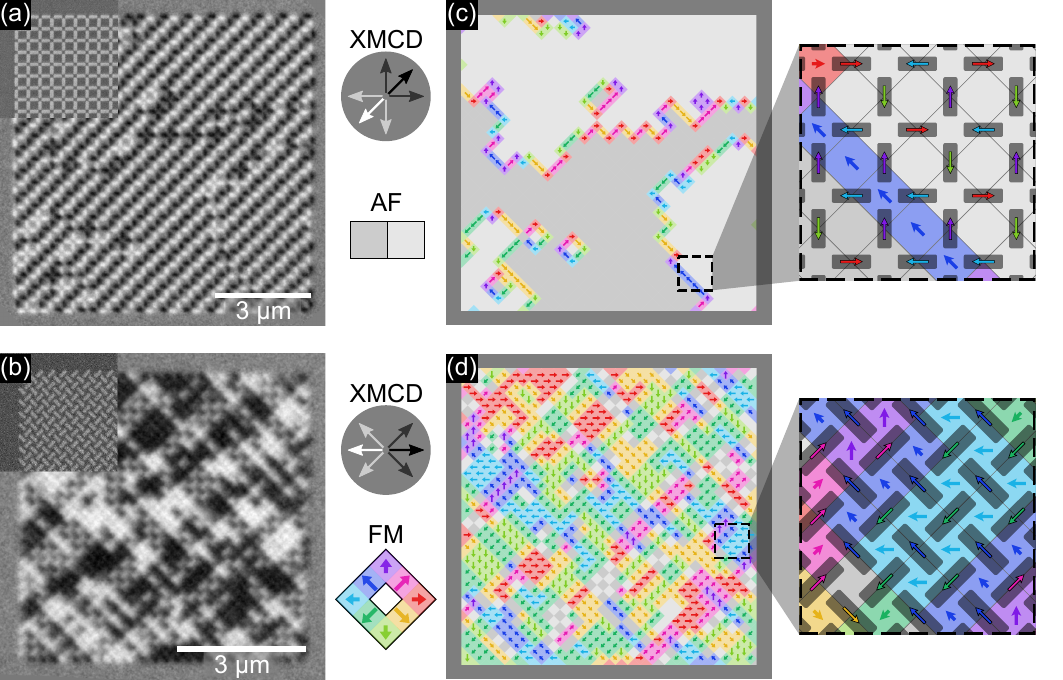}
    \caption{Relaxed states of a square ($\theta = \SI{0}{\degree}$) and pinwheel ($\theta = \SI{45}{\degree}$) ASI. 
    (a) and (b) XMCD-PEEM images, with top left corner insets of a SEM image showing the rotation and shape of the nanomagnets.
    (c) and (d) Analyzed magnetization maps. 
    Gray areas indicate the two possible phases of antiferromagnetically ordered regions.
    Colored regions are ferromagnetically ordered with a net magnetization as depicted in the legend. 
    On the right, the relation between nanomagnet states and cell magnetization is illustrated. 
    The length of the cell arrows correspond to the magnitude of the cell magnetization. }
    \label{fig:peem-decoded}
\end{figure*}

For this study, ensembles of $25\times25$ nanomagnets with lateral dimensions $l\times w=\qtyproduct{220x80}{\nano\meter}$ and thickness $t=\SI{3}{\nano\meter}$ were patterned using electron beam lithography (EBL) and lift-off. 
Each sample featured arrays with a range of different element rotations, $\theta$, corresponding to the schematics in \cref{fig:geometry}.
The magnetic material was deposited by e-beam evaporation of permalloy (\ce{Ni80Fe20}), followed by a \SI{2}{\nano\meter} \ce{Al} layer serving as an oxidation barrier.
The thickness of the magnetic film was chosen so that the ensembles could be easily thermalized \textit{in situ}. 
Moreover, it is imperative to keep the protective Al oxide layer thin so as to permit a strong photoemission signal from the buried magnetic layer, i.e., even if this comes at the expense of partial oxidation of the permalloy.

\begin{figure*}[h!t]
    \centering
    \includegraphics[width=\textwidth]{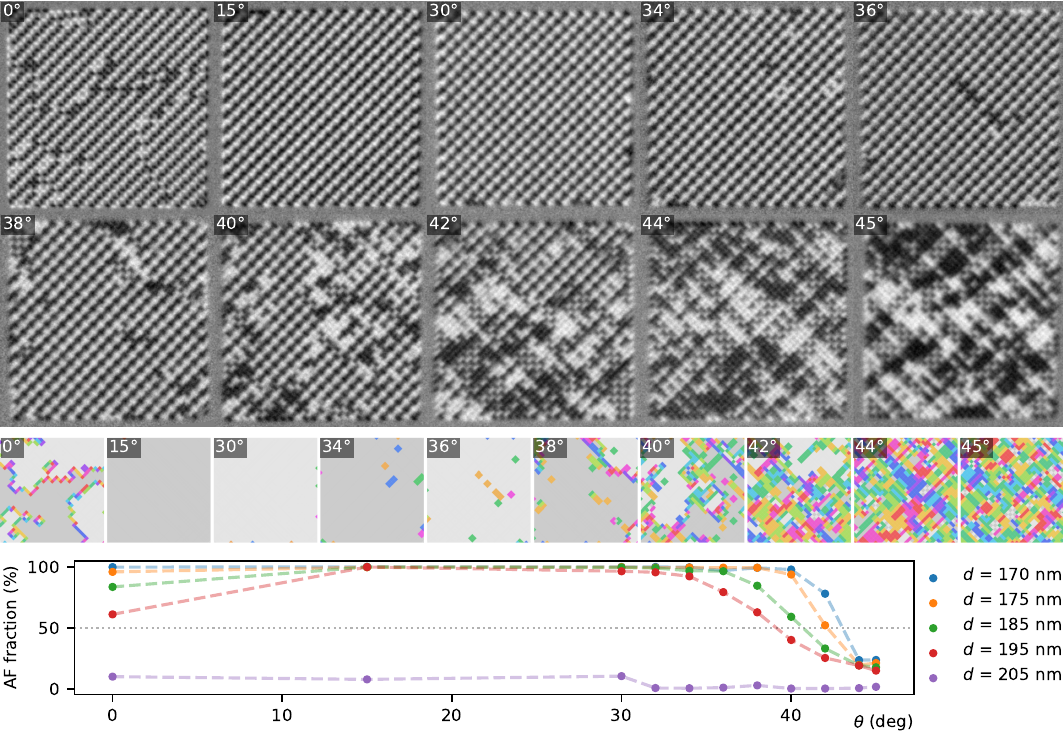}
    \caption{Measured fraction of antiferromagnetic Voronoi cells  for the experimental ensembles. 
    Five different separations $d$, kept constant across the rotation series from \SIrange{0}{45}{\degree}, were investigated.
    XMCD-PEEM images (top panel) show the PEEM images of the relaxed states for the $d=\SI{185}{\nano\meter}$ series.
    Below the PEEM images are graphical representations of the Voronoi cell magnetization (cf. \cref{fig:peem-decoded}).
    The graph shows the fraction of AF Voronoi cells for all separations $d$ as a function of $\theta$.
    }
    \label{fig:af-fraction}
\end{figure*}

Rotation of the elements in this ASI system (see \cref{fig:geometry}(a)) alters the effective coupling between the nanomagnets due to the concentrated stray fields at their ends.
Increasing the rotation angle $\theta$ on a fixed lattice will effectively decouple the elements\citep{Massouras2020}.
In this study, we reduce the lattice constant with increasing rotation angle $\theta$, in order to retain a strong interisland coupling, see \cref{fig:geometry}(b).
We keep the parameter $d$, shown in \cref{fig:geometry}(c), fixed for all element rotations.%

Modeling the AF--FM transition of a full array presents a computational challenge because of the large differences in scale between the internal magnetization of a single nanomagnet to the macroscopic scale of the nanomagnet array.
In order to establish an accurate analytical stray-field model, the stray field of an isolated single nanomagnet is calculated using a micromagnetic simulator\cite{Vansteenkiste2014,Leliaert2018,Beg2022}.
However, this micromagnetic approach is not feasible for the interactions of the full ensemble, due to the high computational cost, and an analytical representation of the stray field for a single nanomagnet is required.
We use the micromagnetic framework $\text{mumax}^3$\citep{Vansteenkiste2014,Leliaert2018} for the micromagnetic calculations and the Ubermag package\citep{Beg2022} for comparison with analytically expressed fields.
The relaxed magnetization in the absence of an external field was computed using micromagnetic simulation, and the stray field was calculated from the resulting magnetization texture.
Each nanomagnet was modeled as a \qtyproduct{220x80}{\nano\meter} rectangle with rounded corners and a thickness of \SI{3}{\nano\meter}.
The saturation magnetization was set at $M_\text{S}=\SI{860}{\kilo\ampere\per\meter}$ and the exchange stiffness at $A_\text{ex} = \SI{1.3e-11}{\joule\per\meter}$\citep{Maicas2008}.
A simulation mesh of $512\times512$ cells was adopted, where each cell had a side length of \SI{2}{\nano\meter}, well below the exchange length $l_\text{ex} = \sqrt{2A_\text{ex}/\mu_0M_\text{S}^2} \approx \SI{5}{\nano\meter}$.

X-ray magnetic circular dichroism photoemission electron microscopy (XMCD-PEEM) was carried out on the PEEM3 beamline at ALS and was used to image the magnetic state of the ensembles. 
Prior to imaging at room temperature, the samples were brought to $T=\SI{385}{\kelvin}$ to thermalize the system, so as to avoid quenched, metastable, high-energy states. 
Complete thermalization was confirmed from the XMCD-PEEM images, showing no magnetic contrast, thus indicating full superparamagnetic behavior on the time scale of imaging (see Suppl. fig. S4).
After thermalization, the sample was brought to room temperature, resulting in a relaxed state governed by the intrinsic dipolar interactions.

Magnetic contrast images of the $\theta = \SI{0}{\degree}$ and $\theta =\SI{45}{\degree}$ ASIs are shown in \cref{fig:peem-decoded}(a) and (b) with the typical AF and FM order, respectively.  
The magnetization of each nanomagnet is identified  from the XMCD-PEEM images, with the help of machine learning, to create a graphical representation of the ensemble state.
In \cref{fig:peem-decoded}(c) and (d), we divide the ASI into so-called Voronoi cells (VCs), where each cell is assigned the net magnetization of four nearest neigbor nanomagnets.
The $\theta = \SI{0}{\degree}$ ASI in \cref{fig:peem-decoded}(c) shows extended regions with zero net magnetic moment, AF-ordered domains divided by domain walls of finite magnetic moment.
In contrast, the $\theta = \SI{45}{\degree}$ ASI in \cref{fig:peem-decoded}(d) displays numerous smaller FM-ordered regions with a net magnetic moment, mimicking the domains of a conventional ferromagnetic material.
As these square and pinwheel ASI ensembles show distinct AF and FM order, respectively, this finding suggests successful relaxation of the arrays and sufficient dipolar magnetic coupling, to study their long-range order.

The main results of the measurement series, where the element rotation $\theta$ is gradually increased while $d$ is kept constant, are shown in \cref{fig:af-fraction}. 
(A complete overview of the XMCD-PEEM images discussed in this study is shown in the Supplemental Material\citep{supplemental}.)
The AF-ordered VCs dominate for all rotation angles below \SI{30}{\degree} irrespective of separation $d$, except for the decoupled $d>\SI{200}{\nano\meter}$ system.
The transition to FM ordering starts at around $\theta = \SI{39}{\degree}$ for the ASI with $d=\SI{195}{\nano\meter}$.
The strongest coupled system, with $d=\SI{170}{\nano\meter}$, exhibits the transition at $\theta = \SI{43}{\degree}$.
The transition angle, $\theta_\text{t}$, is seen to decrease monotonically with reduced magnetic coupling.
We note that the observed transition angles differ significantly from the $\theta_\text{t} \approx \SI{35}{\degree}$ reported in previous work\cite{Macedo2018,Macauley2020} where a simple point-dipole model was used.

\begin{figure*}[h!t]
    \centering
    \includegraphics[width=\textwidth]{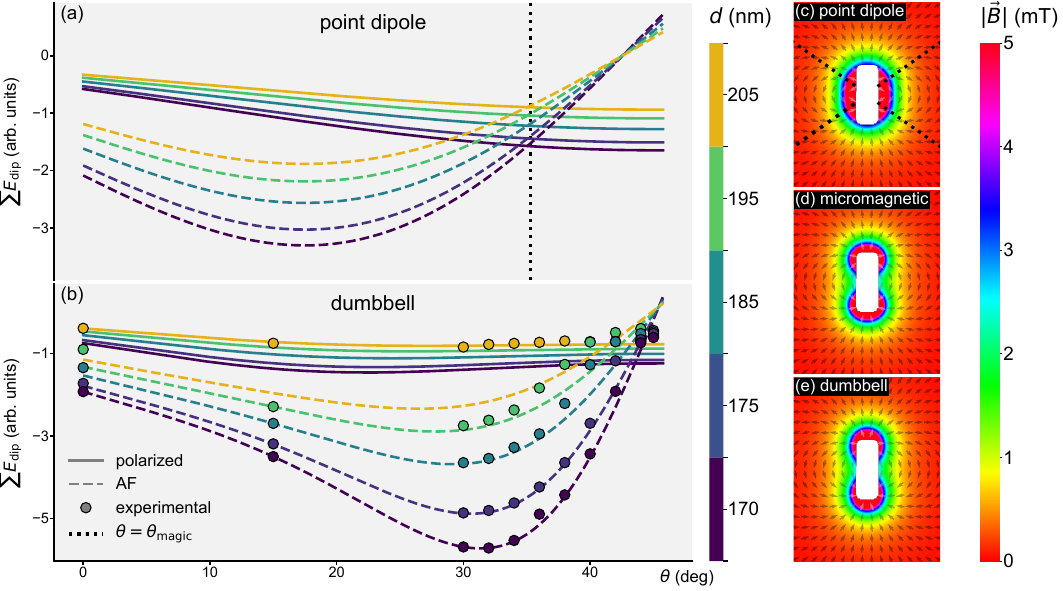}
    \caption{Ensemble state energies and nanomagnet stray fields. 
    Two example states are chosen when calculating the energies, a $100\%$ AF-ordered state and a polarized state, i.e., a $100\%$ FM-ordered state.
    (a) energies for the polarized and AF states, as calculated by the point-dipole model.
    (b) energies for the polarized, AF, and experimental states, as calculated by the dumbbell-dipole model. (c), (d), and (e) show the calculated stray fields for a single \qtyproduct{220x80x3}{\nano\meter} nanomagnet (white rectangle), for the point-dipole model, the micromagnetic simulation, and the dumbbell-dipole model, respectively. 
    The field strength color bar is truncated at \SI{5}{\milli\tesla}. 
    }
    \label{fig:energies-fields}
\end{figure*}

Imperfect AF ordering is observed for $\theta = \SI{0}{\degree}$. 
For this element rotation, we find a multidomain AF ordering with FM ordered domain walls, which may be explained by reduced coupling across the nanomagnet corners.

In the AF-ordered square ASI, the next nearest parallel oriented nanomagnets feature antiparallel spin alignment. 
The change in long-range order from AF to FM with $\theta$ increasing from $\SI{0}{\degree}\rightarrow\SI{45}{\degree}$, implies that for $\theta > \theta_\text{t}$ no neighbors align antiparallel. 
Specifically, the antiparallel magnetization alignment of the next nearest neighbors (a defining feature of AF order in the square ASI) will flip when their stray-field interaction energy favors parallel spin alignment.
With this assumption, we can estimate the transition angle of an ensemble, $\theta_\text{t}$, based on an analytical expression for the stray field.

With the ASI nanomagnets represented as point dipoles, the stray field is described by
\begin{equation}
    \vec{H}_{\text{point dipole}}(\vec{r}) = \frac{1}{4\pi} \left[\frac{3(\vec{m}_{\text{dip}} \cdot \vec{r}) \vec{r}}{r^5} - \frac{\vec{m}_{\text{dip}}}{r^3} \right],
    \label{eq:point-dip}
\end{equation}
where $\vec{m}_{\text{dip}}$ is the magnetic moment of a single nanomagnet and $\vec{r}$ is the position vector from the point dipole.
The preferred orientation of two interacting point dipoles -- parallel or antiparallel -- depends on the angle $\phi$ between the magnetization vector $\vec{m}_{\text{dip}}$ of one dipole and the relative position vector $\vec{r}$ to the other.
Two dipoles positioned head-to-tail favor FM ordering, while two magnets placed side-by-side favor AF ordering.
This change in favored magnetic ordering is due to a switch of sign in the component of the stray field parallel to the dipole magnetization ($\vec{m}_{\text{dip}}$), $H_\parallel$. 
Solving \cref{eq:point-dip} for $H_\parallel = 0$, one finds that this reversal takes place at an angle $\phi_\text{t} = \arctan(1/\sqrt{2}) \approx \SI{35.3}{\degree}$ (cf. \cref{fig:energies-fields}c).
Interestingly, this angle is the same regardless of the distance $r$ and is commonly referred to as the `magic angle'\cite{Erickson1993,deSousa2004}.

When modeling the full ensembles in the point-dipole approximation (using flatspin\citep{Flatspin2022}), the AF--FM transition takes place at element rotation angle $\theta\approx\SI{35.1}{\degree}$, coinciding with the magic angle, independent of the interisland coupling distance $d$. 
At this angle, the nature of the interaction energy for parallel neighbors changes.
The minimum energy state shifts from ``two aligned--two anti-aligned'' to ``all aligned'' macrospins, as a result of the asymmetry between the $x$ and $y$ components of the point-dipole field (\cref{eq:point-dip}).
This angle was cast as the transition angle for the transition between AF and FM order in previous work relying on Monte-Carlo simulations within the point-dipole model\cite{Macedo2018,Macauley2020,Massouras2020}.
However, the point-dipole approximation misses key features of the actual stray-field coupling as it fails to capture the observed variation in transition angle $\theta_\text{t}$ with nanomagnet separation.

The stray fields of the point-dipole model and the micromagnetic simulation are shown in \cref{fig:energies-fields}(c) and (d), respectively.
By comparison with the micromagnetic stray field, we find that a dumbbell-dipole model\citep{Castelnovo2008,Moller2009} captures the characteristics of the nanomagnet stray field better than the point-dipole model (see \cref{fig:energies-fields}(e)).
Finite size nanomagnets in artificial spin ice were previously  modeled in a dumbbell-dipole-like manner to obtain vertex energies in a Monte Carlo simulation\citep{Moller2006,Moller2009}.
In our work, we model the stray field of our finite size nanomagnets as the stray field from a dumbbell dipole.

The dumbbell model emulates two magnetic charges, $\pm q_\text{m}$, separated by a distance $d_\text{dip}$ on the long-axis of the magnet.
The magnitude of the total magnetic moment of a nanomagnet, $\lvert{\vec{m}_\text{dip}}\rvert$, is given by $M_\text{S}\cdot V$, where $M_\text{S}$ represents the saturation magnetization of the magnetic material comprising the nanomagnet and $V$ is the nanomagnet volume.
The effective stray field of the single magnet at a position $\vec{r}$ relative to its center in this model is given by 
\begin{align}
    \vec{H}_\text{dumbbell}(\vec{r}) &= \vec{H}_\text{monopole}(\vec{r}-\vec{a}) - \vec{H}_\text{monopole}(\vec{r}+\vec{a}) ,\nonumber\\ 
    &= \frac{q_m}{4\pi} \left[ \frac{\vec{r} - \vec{a}}{\left| \vec{r} - \vec{a} \right|^3} - \frac{\vec{r} + \vec{a}}{\left| \vec{r} + \vec{a} \right|^3} \right],
    \label{eq:dumbbell}
\end{align}
where $\vec{a}=d_\text{dip}\cdot\hat{d}_\text{dip}/2$ is the distance vector separating the center of the magnet from its positive end. 
For the systems considered in this letter, we find that a naive, unoptimized dumbbell model of monopoles with $d_\text{dip} = l$ and $q_\text{m} \cdot d_\text{dip} = \lvert{\vec{m}_\text{dip}}\rvert$ is sufficient to match the experimental data. 

The magnetic long-range order is evaluated by considering the total energies for the investigated ASI rotation angles $\theta = \SI{0}{\degree}\rightarrow\SI{45}{\degree}$ within the scope of the dumbbell-dipole model.
We find AF and FM long-range order for angles close to \SI{0}{\degree} and \SI{45}{\degree}, respectively. 
Taking the transition angle $\theta_\text{t}$ as the pinwheel rotation angle where the total energy of a FM state drops below the energy of the AF state, the energies of the ensemble for the point-dipole and dumbbell-dipole stray-field models, respectively, are shown in \cref{fig:energies-fields}a-b.
The state energy is calculated by summation over all interactions for each of the experimentally observed states (circular data points), as well as for the fully polarized and AF-ordered states.
The estimated transition angles from simulations and experimental results are compared in \cref{tab:angles}, with the simulated transition angles based on the FM and AF energy calculations.
The measured angles are those where the AF fraction is interpolated at $50\%$.

\begin{table}
\centering
\begin{tabular}{|c|c|c|c|}
\hline
$d$ (nm) & \multicolumn{2}{c|}{$E_\text{AF}=E_\text{pol}$} & 50\% AF \\
 & point dipole & dumbbell & measured \\ \hline
170 & \SI{35.1}{\degree} & \SI{43.5}{\degree} & \SI{43.0}{\degree} \\
175 & \SI{35.1}{\degree} & \SI{43.3}{\degree} & \SI{42.1}{\degree} \\
185 & \SI{35.1}{\degree} & \SI{42.7}{\degree} & \SI{40.7}{\degree} \\
195 & \SI{35.1}{\degree} & \SI{42.0}{\degree} & \SI{39.1}{\degree} \\
205 & \SI{35.1}{\degree} & \SI{41.4}{\degree} & N/A \\
\hline
\end{tabular}
\caption{Transition angles $\theta_\text{t}$, where the calculated energies of the antiferromagnetic state ($E_\text{AF}$) is equal to the energy of the polarized state ($E_\text{pol}$), for point-dipole and dumbbell-dipole models at different $d$ values. 
Also included: experimental transition angle taken to be the angle where the measured fraction of AF Voronoi cells is $50\%$ (by interpolation).
}
\label{tab:angles}
\end{table}

From the stray-field maps in \cref{fig:energies-fields}c-e, it is clear that the dumbbell model better captures the characteristic anisotropy and magnitude of the calculated stray fields of the micromagnetically modeled nanomagnet, in particular in the near-field region depicted here.  
Moreover, dumbbell-dipole model accurately captures the experimental observations with a transition angle $\theta_\text{t}$ dependent on the intermagnet coupling.
The AF--FM transition is observed for angles ranging from \SIrange{39}{43}{\degree}, while simulations using this model indicate an AF--FM energy crossover for angles from \SI{41.5}{\degree} to \SI{43.3}{\degree}. 
This close correspondence underscores the advantage of using the dumbbell-dipole model for an accurate representation of the physical behavior of the system.

A notable exception to the AF order, observed at small angles $\theta$, is found for the $d = \SI{205}{\nano\meter}$-series in \cref{fig:af-fraction}, where no experimental transition is found.
The ensembles with this nanomagnet separation are all nearly polarized in the same direction.
This finding suggests that dipolar coupling in this particular ensemble is too small to drive long-range order.
The observed polarization may be due to a small stray field present during annealing inside the PEEM-3 microscope. 

In conclusion, this study shows a transition from distinct AF order to pronounced FM order in a square-to-pinwheel artificial spin ice system.
We find that the transition angle is dependent on the interisland coupling, a behavior not captured by the conventional point-dipole approximation.
By introducing a dumbbell-dipole model, we achieve excellent agreement with experimental observations and micromagnetic simulations of the near-field interactions.
We maintain that this model offers a suitable framework for simulation of a variety of nanomagnet shapes and textures.
Our approach improves modeling of magnetic ordering in ASIs, which is key to fundamental research as well as technological applications.

\vspace*{5mm}
\begin{acknowledgments}
This research used resources of the Advanced Light Source, which is a DOE Office of Science User Facilities under contract no. DE-AC02-05CH11231. 

The Research Council of Norway is acknowledged for the support to the Norwegian Micro- and Nanofabrication Facility, NorFab, project no. 295864. The samples were fabricated at NTNU NanoLab. 

This work was funded by the Norwegian Research Council through the IKTPLUSS project SOCRATES (Grant no. 270961) and the TEKNOKONVERGENS project SPrINTER (Grant No. 331821), and by the EU FET-Open RIA project SpinENGINE (Grant no. 861618).

\end{acknowledgments}
\bibliography{bibliography}%

\begin{thebibliography}{25}%
\makeatletter
\providecommand \@ifxundefined [1]{%
 \@ifx{#1\undefined}
}%
\providecommand \@ifnum [1]{%
 \ifnum #1\expandafter \@firstoftwo
 \else \expandafter \@secondoftwo
 \fi
}%
\providecommand \@ifx [1]{%
 \ifx #1\expandafter \@firstoftwo
 \else \expandafter \@secondoftwo
 \fi
}%
\providecommand \natexlab [1]{#1}%
\providecommand \enquote  [1]{``#1''}%
\providecommand \bibnamefont  [1]{#1}%
\providecommand \bibfnamefont [1]{#1}%
\providecommand \citenamefont [1]{#1}%
\providecommand \href@noop [0]{\@secondoftwo}%
\providecommand \href [0]{\begingroup \@sanitize@url \@href}%
\providecommand \@href[1]{\@@startlink{#1}\@@href}%
\providecommand \@@href[1]{\endgroup#1\@@endlink}%
\providecommand \@sanitize@url [0]{\catcode `\\12\catcode `\$12\catcode
  `\&12\catcode `\#12\catcode `\^12\catcode `\_12\catcode `\%12\relax}%
\providecommand \@@startlink[1]{}%
\providecommand \@@endlink[0]{}%
\providecommand \url  [0]{\begingroup\@sanitize@url \@url }%
\providecommand \@url [1]{\endgroup\@href {#1}{\urlprefix }}%
\providecommand \urlprefix  [0]{URL }%
\providecommand \Eprint [0]{\href }%
\providecommand \doibase [0]{https://doi.org/}%
\providecommand \selectlanguage [0]{\@gobble}%
\providecommand \bibinfo  [0]{\@secondoftwo}%
\providecommand \bibfield  [0]{\@secondoftwo}%
\providecommand \translation [1]{[#1]}%
\providecommand \BibitemOpen [0]{}%
\providecommand \bibitemStop [0]{}%
\providecommand \bibitemNoStop [0]{.\EOS\space}%
\providecommand \EOS [0]{\spacefactor3000\relax}%
\providecommand \BibitemShut  [1]{\csname bibitem#1\endcsname}%
\let\auto@bib@innerbib\@empty
\bibitem [{\citenamefont {Wang}\ \emph {et~al.}(2006)\citenamefont {Wang},
  \citenamefont {Nisoli}, \citenamefont {Freitas}, \citenamefont {Li},
  \citenamefont {McConville}, \citenamefont {Cooley}, \citenamefont {Lund},
  \citenamefont {Samarth}, \citenamefont {Leighton}, \citenamefont {Crespi},\
  and\ \citenamefont {Schiffer}}]{Wang2006}%
  \BibitemOpen
  \bibfield  {author} {\bibinfo {author} {\bibfnamefont {R.~F.}\ \bibnamefont
  {Wang}}, \bibinfo {author} {\bibfnamefont {C.}~\bibnamefont {Nisoli}},
  \bibinfo {author} {\bibfnamefont {R.~S.}\ \bibnamefont {Freitas}}, \bibinfo
  {author} {\bibfnamefont {J.}~\bibnamefont {Li}}, \bibinfo {author}
  {\bibfnamefont {W.}~\bibnamefont {McConville}}, \bibinfo {author}
  {\bibfnamefont {B.~J.}\ \bibnamefont {Cooley}}, \bibinfo {author}
  {\bibfnamefont {M.~S.}\ \bibnamefont {Lund}}, \bibinfo {author}
  {\bibfnamefont {N.}~\bibnamefont {Samarth}}, \bibinfo {author} {\bibfnamefont
  {C.}~\bibnamefont {Leighton}}, \bibinfo {author} {\bibfnamefont {V.~H.}\
  \bibnamefont {Crespi}},\ and\ \bibinfo {author} {\bibfnamefont
  {P.}~\bibnamefont {Schiffer}},\ }\bibfield  {title} {\bibinfo {title}
  {Artificial `spin ice' in a geometrically frustrated lattice of nanoscale
  ferromagnetic islands},\ }\href {https://doi.org/10.1038/nature04447}
  {\bibfield  {journal} {\bibinfo  {journal} {Nature}\ }\textbf {\bibinfo
  {volume} {439}},\ \bibinfo {pages} {303} (\bibinfo {year}
  {2006})}\BibitemShut {NoStop}%
\bibitem [{\citenamefont {Mengotti}\ \emph {et~al.}(2011)\citenamefont
  {Mengotti}, \citenamefont {Heyderman}, \citenamefont {Rodr{\'i}guez},
  \citenamefont {Nolting}, \citenamefont {H{\"u}gli},\ and\ \citenamefont
  {Braun}}]{Mengotti2011}%
  \BibitemOpen
  \bibfield  {author} {\bibinfo {author} {\bibfnamefont {E.}~\bibnamefont
  {Mengotti}}, \bibinfo {author} {\bibfnamefont {L.~J.}\ \bibnamefont
  {Heyderman}}, \bibinfo {author} {\bibfnamefont {A.~F.}\ \bibnamefont
  {Rodr{\'i}guez}}, \bibinfo {author} {\bibfnamefont {F.}~\bibnamefont
  {Nolting}}, \bibinfo {author} {\bibfnamefont {R.~V.}\ \bibnamefont
  {H{\"u}gli}},\ and\ \bibinfo {author} {\bibfnamefont {H.-B.}\ \bibnamefont
  {Braun}},\ }\bibfield  {title} {\bibinfo {title} {Real-space observation of
  emergent magnetic monopoles and associated dirac strings in artificial kagome
  spin ice},\ }\href {https://doi.org/10.1038/nphys1794} {\bibfield  {journal}
  {\bibinfo  {journal} {Nature Physics}\ }\textbf {\bibinfo {volume} {7}},\
  \bibinfo {pages} {68} (\bibinfo {year} {2011})}\BibitemShut {NoStop}%
\bibitem [{\citenamefont {Skj{\ae}rv{\o}}\ \emph {et~al.}(2020)\citenamefont
  {Skj{\ae}rv{\o}}, \citenamefont {Marrows}, \citenamefont {Stamps},\ and\
  \citenamefont {Heyderman}}]{Skjaervo2020}%
  \BibitemOpen
  \bibfield  {author} {\bibinfo {author} {\bibfnamefont {S.~H.}\ \bibnamefont
  {Skj{\ae}rv{\o}}}, \bibinfo {author} {\bibfnamefont {C.~H.}\ \bibnamefont
  {Marrows}}, \bibinfo {author} {\bibfnamefont {R.~L.}\ \bibnamefont
  {Stamps}},\ and\ \bibinfo {author} {\bibfnamefont {L.~J.}\ \bibnamefont
  {Heyderman}},\ }\bibfield  {title} {\bibinfo {title} {Advances in artificial
  spin ice},\ }\href {https://doi.org/10.1038/s42254-019-0118-3} {\bibfield
  {journal} {\bibinfo  {journal} {Nature Reviews Physics}\ }\textbf {\bibinfo
  {volume} {2}},\ \bibinfo {pages} {13} (\bibinfo {year} {2020})}\BibitemShut
  {NoStop}%
\bibitem [{\citenamefont {May}\ \emph {et~al.}(2021)\citenamefont {May},
  \citenamefont {Saccone}, \citenamefont {{van den Berg}}, \citenamefont
  {Askey}, \citenamefont {Hunt},\ and\ \citenamefont {Ladak}}]{May2021}%
  \BibitemOpen
  \bibfield  {author} {\bibinfo {author} {\bibfnamefont {A.}~\bibnamefont
  {May}}, \bibinfo {author} {\bibfnamefont {M.}~\bibnamefont {Saccone}},
  \bibinfo {author} {\bibfnamefont {A.}~\bibnamefont {{van den Berg}}},
  \bibinfo {author} {\bibfnamefont {J.}~\bibnamefont {Askey}}, \bibinfo
  {author} {\bibfnamefont {M.}~\bibnamefont {Hunt}},\ and\ \bibinfo {author}
  {\bibfnamefont {S.}~\bibnamefont {Ladak}},\ }\bibfield  {title} {\bibinfo
  {title} {Magnetic charge propagation upon a {{3D}} artificial spin-ice},\
  }\href {https://doi.org/10.1038/s41467-021-23480-7} {\bibfield  {journal}
  {\bibinfo  {journal} {Nature Communications}\ }\textbf {\bibinfo {volume}
  {12}},\ \bibinfo {pages} {3217} (\bibinfo {year} {2021})}\BibitemShut
  {NoStop}%
\bibitem [{\citenamefont {Digernes}\ \emph {et~al.}(2021)\citenamefont
  {Digernes}, \citenamefont {Str{\o}mberg}, \citenamefont {Vaz}, \citenamefont
  {Kleibert}, \citenamefont {Grepstad},\ and\ \citenamefont
  {Folven}}]{Digernes2021}%
  \BibitemOpen
  \bibfield  {author} {\bibinfo {author} {\bibfnamefont {E.}~\bibnamefont
  {Digernes}}, \bibinfo {author} {\bibfnamefont {A.}~\bibnamefont
  {Str{\o}mberg}}, \bibinfo {author} {\bibfnamefont {C.~A.~F.}\ \bibnamefont
  {Vaz}}, \bibinfo {author} {\bibfnamefont {A.}~\bibnamefont {Kleibert}},
  \bibinfo {author} {\bibfnamefont {J.~K.}\ \bibnamefont {Grepstad}},\ and\
  \bibinfo {author} {\bibfnamefont {E.}~\bibnamefont {Folven}},\ }\bibfield
  {title} {\bibinfo {title} {Anisotropy and domain formation in a dipolar
  magnetic metamaterial},\ }\href {https://doi.org/10.1063/5.0045450}
  {\bibfield  {journal} {\bibinfo  {journal} {Applied Physics Letters}\
  }\textbf {\bibinfo {volume} {118}},\ \bibinfo {pages} {202404} (\bibinfo
  {year} {2021})}\BibitemShut {NoStop}%
\bibitem [{\citenamefont {Jensen}\ \emph {et~al.}(2018)\citenamefont {Jensen},
  \citenamefont {Folven},\ and\ \citenamefont {Tufte}}]{Jensen2018}%
  \BibitemOpen
  \bibfield  {author} {\bibinfo {author} {\bibfnamefont {J.~H.}\ \bibnamefont
  {Jensen}}, \bibinfo {author} {\bibfnamefont {E.}~\bibnamefont {Folven}},\
  and\ \bibinfo {author} {\bibfnamefont {G.}~\bibnamefont {Tufte}},\ }\bibfield
   {title} {\bibinfo {title} {Computation in artificial spin ice},\ }in\ \href
  {https://doi.org/10.1162/isal_a_00011} {\emph {\bibinfo {booktitle}
  {{{ALIFE}} 2018: {{The}} 2018 {{Conference}} on {{Artificial Life}}}}}\
  (\bibinfo  {publisher} {MIT Press},\ \bibinfo {address} {Tokyo, Japan},\
  \bibinfo {year} {2018})\ pp.\ \bibinfo {pages} {15--22}\BibitemShut {NoStop}%
\bibitem [{\citenamefont {Gartside}\ \emph {et~al.}(2022)\citenamefont
  {Gartside}, \citenamefont {Stenning}, \citenamefont {Vanstone}, \citenamefont
  {Holder}, \citenamefont {Arroo}, \citenamefont {Dion}, \citenamefont
  {Caravelli}, \citenamefont {Kurebayashi},\ and\ \citenamefont
  {Branford}}]{Gartside2022}%
  \BibitemOpen
  \bibfield  {author} {\bibinfo {author} {\bibfnamefont {J.~C.}\ \bibnamefont
  {Gartside}}, \bibinfo {author} {\bibfnamefont {K.~D.}\ \bibnamefont
  {Stenning}}, \bibinfo {author} {\bibfnamefont {A.}~\bibnamefont {Vanstone}},
  \bibinfo {author} {\bibfnamefont {H.~H.}\ \bibnamefont {Holder}}, \bibinfo
  {author} {\bibfnamefont {D.~M.}\ \bibnamefont {Arroo}}, \bibinfo {author}
  {\bibfnamefont {T.}~\bibnamefont {Dion}}, \bibinfo {author} {\bibfnamefont
  {F.}~\bibnamefont {Caravelli}}, \bibinfo {author} {\bibfnamefont
  {H.}~\bibnamefont {Kurebayashi}},\ and\ \bibinfo {author} {\bibfnamefont
  {W.~R.}\ \bibnamefont {Branford}},\ }\bibfield  {title} {\bibinfo {title}
  {Reconfigurable training and reservoir computing in an artificial spin-vortex
  ice via spin-wave fingerprinting},\ }\href
  {https://doi.org/10.1038/s41565-022-01091-7} {\bibfield  {journal} {\bibinfo
  {journal} {Nature Nanotechnology}\ }\textbf {\bibinfo {volume} {17}},\
  \bibinfo {pages} {460} (\bibinfo {year} {2022})}\BibitemShut {NoStop}%
\bibitem [{\citenamefont {Digernes}\ \emph {et~al.}(2020)\citenamefont
  {Digernes}, \citenamefont {Sl{\"o}etjes}, \citenamefont {Str{\o}mberg},
  \citenamefont {Bang}, \citenamefont {Olsen}, \citenamefont {Arenholz},
  \citenamefont {Chopdekar}, \citenamefont {Grepstad},\ and\ \citenamefont
  {Folven}}]{Digernes2020}%
  \BibitemOpen
  \bibfield  {author} {\bibinfo {author} {\bibfnamefont {E.}~\bibnamefont
  {Digernes}}, \bibinfo {author} {\bibfnamefont {S.~D.}\ \bibnamefont
  {Sl{\"o}etjes}}, \bibinfo {author} {\bibfnamefont {A.}~\bibnamefont
  {Str{\o}mberg}}, \bibinfo {author} {\bibfnamefont {A.~D.}\ \bibnamefont
  {Bang}}, \bibinfo {author} {\bibfnamefont {F.~K.}\ \bibnamefont {Olsen}},
  \bibinfo {author} {\bibfnamefont {E.}~\bibnamefont {Arenholz}}, \bibinfo
  {author} {\bibfnamefont {R.~V.}\ \bibnamefont {Chopdekar}}, \bibinfo {author}
  {\bibfnamefont {J.~K.}\ \bibnamefont {Grepstad}},\ and\ \bibinfo {author}
  {\bibfnamefont {E.}~\bibnamefont {Folven}},\ }\bibfield  {title} {\bibinfo
  {title} {Direct imaging of long-range ferromagnetic and antiferromagnetic
  order in a dipolar metamaterial},\ }\href
  {https://doi.org/10.1103/PhysRevResearch.2.013222} {\bibfield  {journal}
  {\bibinfo  {journal} {Physical Review Research}\ }\textbf {\bibinfo {volume}
  {2}},\ \bibinfo {pages} {013222} (\bibinfo {year} {2020})}\BibitemShut
  {NoStop}%
\bibitem [{\citenamefont {Mac{\^e}do}\ \emph {et~al.}(2018)\citenamefont
  {Mac{\^e}do}, \citenamefont {Macauley}, \citenamefont {Nascimento},\ and\
  \citenamefont {Stamps}}]{Macedo2018}%
  \BibitemOpen
  \bibfield  {author} {\bibinfo {author} {\bibfnamefont {R.}~\bibnamefont
  {Mac{\^e}do}}, \bibinfo {author} {\bibfnamefont {G.~M.}\ \bibnamefont
  {Macauley}}, \bibinfo {author} {\bibfnamefont {F.~S.}\ \bibnamefont
  {Nascimento}},\ and\ \bibinfo {author} {\bibfnamefont {R.~L.}\ \bibnamefont
  {Stamps}},\ }\bibfield  {title} {\bibinfo {title} {Apparent ferromagnetism in
  the pinwheel artificial spin ice},\ }\href
  {https://doi.org/10.1103/PhysRevB.98.014437} {\bibfield  {journal} {\bibinfo
  {journal} {Physical Review B}\ }\textbf {\bibinfo {volume} {98}},\ \bibinfo
  {pages} {014437} (\bibinfo {year} {2018})}\BibitemShut {NoStop}%
\bibitem [{\citenamefont {Gliga}\ \emph {et~al.}(2017)\citenamefont {Gliga},
  \citenamefont {Hrkac}, \citenamefont {Donnelly}, \citenamefont {B{\"u}chi},
  \citenamefont {Kleibert}, \citenamefont {Cui}, \citenamefont {Farhan},
  \citenamefont {Kirk}, \citenamefont {Chopdekar}, \citenamefont {Masaki},
  \citenamefont {Bingham}, \citenamefont {Scholl}, \citenamefont {Stamps},\
  and\ \citenamefont {Heyderman}}]{Gliga2017b}%
  \BibitemOpen
  \bibfield  {author} {\bibinfo {author} {\bibfnamefont {S.}~\bibnamefont
  {Gliga}}, \bibinfo {author} {\bibfnamefont {G.}~\bibnamefont {Hrkac}},
  \bibinfo {author} {\bibfnamefont {C.}~\bibnamefont {Donnelly}}, \bibinfo
  {author} {\bibfnamefont {J.}~\bibnamefont {B{\"u}chi}}, \bibinfo {author}
  {\bibfnamefont {A.}~\bibnamefont {Kleibert}}, \bibinfo {author}
  {\bibfnamefont {J.}~\bibnamefont {Cui}}, \bibinfo {author} {\bibfnamefont
  {A.}~\bibnamefont {Farhan}}, \bibinfo {author} {\bibfnamefont
  {E.}~\bibnamefont {Kirk}}, \bibinfo {author} {\bibfnamefont {R.~V.}\
  \bibnamefont {Chopdekar}}, \bibinfo {author} {\bibfnamefont {Y.}~\bibnamefont
  {Masaki}}, \bibinfo {author} {\bibfnamefont {N.~S.}\ \bibnamefont {Bingham}},
  \bibinfo {author} {\bibfnamefont {A.}~\bibnamefont {Scholl}}, \bibinfo
  {author} {\bibfnamefont {R.~L.}\ \bibnamefont {Stamps}},\ and\ \bibinfo
  {author} {\bibfnamefont {L.~J.}\ \bibnamefont {Heyderman}},\ }\bibfield
  {title} {\bibinfo {title} {Emergent dynamic chirality in a thermally driven
  artificial spin ratchet},\ }\href {https://doi.org/10.1038/nmat5007}
  {\bibfield  {journal} {\bibinfo  {journal} {Nature Materials}\ }\textbf
  {\bibinfo {volume} {16}},\ \bibinfo {pages} {1106} (\bibinfo {year}
  {2017})}\BibitemShut {NoStop}%
\bibitem [{\citenamefont {Li}\ \emph {et~al.}(2019)\citenamefont {Li},
  \citenamefont {Paterson}, \citenamefont {Macauley}, \citenamefont
  {Nascimento}, \citenamefont {Ferguson}, \citenamefont {Morley}, \citenamefont
  {Rosamond}, \citenamefont {Linfield}, \citenamefont {MacLaren}, \citenamefont
  {Mac{\^e}do}, \citenamefont {Marrows}, \citenamefont {McVitie},\ and\
  \citenamefont {Stamps}}]{Li2019}%
  \BibitemOpen
  \bibfield  {author} {\bibinfo {author} {\bibfnamefont {Y.}~\bibnamefont
  {Li}}, \bibinfo {author} {\bibfnamefont {G.~W.}\ \bibnamefont {Paterson}},
  \bibinfo {author} {\bibfnamefont {G.~M.}\ \bibnamefont {Macauley}}, \bibinfo
  {author} {\bibfnamefont {F.~S.}\ \bibnamefont {Nascimento}}, \bibinfo
  {author} {\bibfnamefont {C.}~\bibnamefont {Ferguson}}, \bibinfo {author}
  {\bibfnamefont {S.~A.}\ \bibnamefont {Morley}}, \bibinfo {author}
  {\bibfnamefont {M.~C.}\ \bibnamefont {Rosamond}}, \bibinfo {author}
  {\bibfnamefont {E.~H.}\ \bibnamefont {Linfield}}, \bibinfo {author}
  {\bibfnamefont {D.~A.}\ \bibnamefont {MacLaren}}, \bibinfo {author}
  {\bibfnamefont {R.}~\bibnamefont {Mac{\^e}do}}, \bibinfo {author}
  {\bibfnamefont {C.~H.}\ \bibnamefont {Marrows}}, \bibinfo {author}
  {\bibfnamefont {S.}~\bibnamefont {McVitie}},\ and\ \bibinfo {author}
  {\bibfnamefont {R.~L.}\ \bibnamefont {Stamps}},\ }\bibfield  {title}
  {\bibinfo {title} {Superferromagnetism and domain-wall topologies in
  artificial ``{{Pinwheel}}'' spin ice},\ }\href
  {https://doi.org/10.1021/acsnano.8b08884} {\bibfield  {journal} {\bibinfo
  {journal} {ACS Nano}\ }\textbf {\bibinfo {volume} {13}},\ \bibinfo {pages}
  {2213} (\bibinfo {year} {2019})}\BibitemShut {NoStop}%
\bibitem [{\citenamefont {Paterson}\ \emph {et~al.}(2019)\citenamefont
  {Paterson}, \citenamefont {Macauley}, \citenamefont {Li}, \citenamefont
  {Mac{\^e}do}, \citenamefont {Ferguson}, \citenamefont {Morley}, \citenamefont
  {Rosamond}, \citenamefont {Linfield}, \citenamefont {Marrows}, \citenamefont
  {Stamps},\ and\ \citenamefont {McVitie}}]{Paterson2019}%
  \BibitemOpen
  \bibfield  {author} {\bibinfo {author} {\bibfnamefont {G.~W.}\ \bibnamefont
  {Paterson}}, \bibinfo {author} {\bibfnamefont {G.~M.}\ \bibnamefont
  {Macauley}}, \bibinfo {author} {\bibfnamefont {Y.}~\bibnamefont {Li}},
  \bibinfo {author} {\bibfnamefont {R.}~\bibnamefont {Mac{\^e}do}}, \bibinfo
  {author} {\bibfnamefont {C.}~\bibnamefont {Ferguson}}, \bibinfo {author}
  {\bibfnamefont {S.~A.}\ \bibnamefont {Morley}}, \bibinfo {author}
  {\bibfnamefont {M.~C.}\ \bibnamefont {Rosamond}}, \bibinfo {author}
  {\bibfnamefont {E.~H.}\ \bibnamefont {Linfield}}, \bibinfo {author}
  {\bibfnamefont {C.~H.}\ \bibnamefont {Marrows}}, \bibinfo {author}
  {\bibfnamefont {R.~L.}\ \bibnamefont {Stamps}},\ and\ \bibinfo {author}
  {\bibfnamefont {S.}~\bibnamefont {McVitie}},\ }\bibfield  {title} {\bibinfo
  {title} {Heisenberg pseudo-exchange and emergent anisotropies in field-driven
  pinwheel artificial spin ice},\ }\href
  {https://doi.org/10.1103/PhysRevB.100.174410} {\bibfield  {journal} {\bibinfo
   {journal} {Physical Review B}\ }\textbf {\bibinfo {volume} {100}},\ \bibinfo
  {pages} {174410} (\bibinfo {year} {2019})}\BibitemShut {NoStop}%
\bibitem [{\citenamefont {Macauley}\ \emph {et~al.}(2020)\citenamefont
  {Macauley}, \citenamefont {Paterson}, \citenamefont {Li}, \citenamefont
  {Mac{\^e}do}, \citenamefont {McVitie},\ and\ \citenamefont
  {Stamps}}]{Macauley2020}%
  \BibitemOpen
  \bibfield  {author} {\bibinfo {author} {\bibfnamefont {G.~M.}\ \bibnamefont
  {Macauley}}, \bibinfo {author} {\bibfnamefont {G.~W.}\ \bibnamefont
  {Paterson}}, \bibinfo {author} {\bibfnamefont {Y.}~\bibnamefont {Li}},
  \bibinfo {author} {\bibfnamefont {R.}~\bibnamefont {Mac{\^e}do}}, \bibinfo
  {author} {\bibfnamefont {S.}~\bibnamefont {McVitie}},\ and\ \bibinfo {author}
  {\bibfnamefont {R.~L.}\ \bibnamefont {Stamps}},\ }\bibfield  {title}
  {\bibinfo {title} {Tuning magnetic order with geometry: Thermalization and
  defects in two-dimensional artificial spin ices},\ }\href
  {https://doi.org/10.1103/PhysRevB.101.144403} {\bibfield  {journal} {\bibinfo
   {journal} {Physical Review B}\ }\textbf {\bibinfo {volume} {101}},\ \bibinfo
  {pages} {144403} (\bibinfo {year} {2020})}\BibitemShut {NoStop}%
\bibitem [{\citenamefont {Massouras}\ \emph {et~al.}(2020)\citenamefont
  {Massouras}, \citenamefont {Lacour}, \citenamefont {Hehn},\ and\
  \citenamefont {Montaigne}}]{Massouras2020}%
  \BibitemOpen
  \bibfield  {author} {\bibinfo {author} {\bibfnamefont {M.}~\bibnamefont
  {Massouras}}, \bibinfo {author} {\bibfnamefont {D.}~\bibnamefont {Lacour}},
  \bibinfo {author} {\bibfnamefont {M.}~\bibnamefont {Hehn}},\ and\ \bibinfo
  {author} {\bibfnamefont {F.}~\bibnamefont {Montaigne}},\ }\bibfield  {title}
  {\bibinfo {title} {Probing the antiferromagnetic-paramagnetic transition in
  artificial spin ice by tuning interactions},\ }\href
  {https://doi.org/10.1103/PhysRevB.101.174421} {\bibfield  {journal} {\bibinfo
   {journal} {Physical Review B}\ }\textbf {\bibinfo {volume} {101}},\ \bibinfo
  {pages} {174421} (\bibinfo {year} {2020})}\BibitemShut {NoStop}%
\bibitem [{\citenamefont {Castelnovo}\ \emph {et~al.}(2008)\citenamefont
  {Castelnovo}, \citenamefont {Moessner},\ and\ \citenamefont
  {Sondhi}}]{Castelnovo2008}%
  \BibitemOpen
  \bibfield  {author} {\bibinfo {author} {\bibfnamefont {C.}~\bibnamefont
  {Castelnovo}}, \bibinfo {author} {\bibfnamefont {R.}~\bibnamefont
  {Moessner}},\ and\ \bibinfo {author} {\bibfnamefont {S.~L.}\ \bibnamefont
  {Sondhi}},\ }\bibfield  {title} {\bibinfo {title} {Magnetic monopoles in spin
  ice},\ }\href {https://doi.org/10.1038/nature06433} {\bibfield  {journal}
  {\bibinfo  {journal} {Nature}\ }\textbf {\bibinfo {volume} {451}},\ \bibinfo
  {pages} {42} (\bibinfo {year} {2008})}\BibitemShut {NoStop}%
\bibitem [{\citenamefont {Vansteenkiste}\ \emph {et~al.}(2014)\citenamefont
  {Vansteenkiste}, \citenamefont {Leliaert}, \citenamefont {Dvornik},
  \citenamefont {Helsen}, \citenamefont {{Garcia-Sanchez}},\ and\ \citenamefont
  {Van~Waeyenberge}}]{Vansteenkiste2014}%
  \BibitemOpen
  \bibfield  {author} {\bibinfo {author} {\bibfnamefont {A.}~\bibnamefont
  {Vansteenkiste}}, \bibinfo {author} {\bibfnamefont {J.}~\bibnamefont
  {Leliaert}}, \bibinfo {author} {\bibfnamefont {M.}~\bibnamefont {Dvornik}},
  \bibinfo {author} {\bibfnamefont {M.}~\bibnamefont {Helsen}}, \bibinfo
  {author} {\bibfnamefont {F.}~\bibnamefont {{Garcia-Sanchez}}},\ and\ \bibinfo
  {author} {\bibfnamefont {B.}~\bibnamefont {Van~Waeyenberge}},\ }\bibfield
  {title} {\bibinfo {title} {The design and verification of {{MuMax3}}},\
  }\href {https://doi.org/10.1063/1.4899186} {\bibfield  {journal} {\bibinfo
  {journal} {AIP Advances}\ }\textbf {\bibinfo {volume} {4}},\ \bibinfo {pages}
  {107133} (\bibinfo {year} {2014})}\BibitemShut {NoStop}%
\bibitem [{\citenamefont {Leliaert}\ \emph {et~al.}(2018)\citenamefont
  {Leliaert}, \citenamefont {Dvornik}, \citenamefont {Mulkers}, \citenamefont
  {Clercq}, \citenamefont {Milo{\v s}evi{\'c}},\ and\ \citenamefont
  {Waeyenberge}}]{Leliaert2018}%
  \BibitemOpen
  \bibfield  {author} {\bibinfo {author} {\bibfnamefont {J.}~\bibnamefont
  {Leliaert}}, \bibinfo {author} {\bibfnamefont {M.}~\bibnamefont {Dvornik}},
  \bibinfo {author} {\bibfnamefont {J.}~\bibnamefont {Mulkers}}, \bibinfo
  {author} {\bibfnamefont {J.~D.}\ \bibnamefont {Clercq}}, \bibinfo {author}
  {\bibfnamefont {M.~V.}\ \bibnamefont {Milo{\v s}evi{\'c}}},\ and\ \bibinfo
  {author} {\bibfnamefont {B.~V.}\ \bibnamefont {Waeyenberge}},\ }\bibfield
  {title} {\bibinfo {title} {Fast micromagnetic simulations on {{GPU}}---recent
  advances made with mumax3},\ }\href
  {https://doi.org/10.1088/1361-6463/aaab1c} {\bibfield  {journal} {\bibinfo
  {journal} {Journal of Physics D: Applied Physics}\ }\textbf {\bibinfo
  {volume} {51}},\ \bibinfo {pages} {123002} (\bibinfo {year}
  {2018})}\BibitemShut {NoStop}%
\bibitem [{\citenamefont {Beg}\ \emph {et~al.}(2022)\citenamefont {Beg},
  \citenamefont {Lang},\ and\ \citenamefont {Fangohr}}]{Beg2022}%
  \BibitemOpen
  \bibfield  {author} {\bibinfo {author} {\bibfnamefont {M.}~\bibnamefont
  {Beg}}, \bibinfo {author} {\bibfnamefont {M.}~\bibnamefont {Lang}},\ and\
  \bibinfo {author} {\bibfnamefont {H.}~\bibnamefont {Fangohr}},\ }\bibfield
  {title} {\bibinfo {title} {Ubermag: {{Toward More Effective Micromagnetic
  Workflows}}},\ }\href {https://doi.org/10.1109/TMAG.2021.3078896} {\bibfield
  {journal} {\bibinfo  {journal} {IEEE Transactions on Magnetics}\ }\textbf
  {\bibinfo {volume} {58}},\ \bibinfo {pages} {1} (\bibinfo {year}
  {2022})}\BibitemShut {NoStop}%
\bibitem [{\citenamefont {Maicas}\ \emph {et~al.}(2008)\citenamefont {Maicas},
  \citenamefont {Ranchal}, \citenamefont {Aroca}, \citenamefont {S{\'a}nchez},\
  and\ \citenamefont {L{\'o}pez}}]{Maicas2008}%
  \BibitemOpen
  \bibfield  {author} {\bibinfo {author} {\bibfnamefont {M.}~\bibnamefont
  {Maicas}}, \bibinfo {author} {\bibfnamefont {R.}~\bibnamefont {Ranchal}},
  \bibinfo {author} {\bibfnamefont {C.}~\bibnamefont {Aroca}}, \bibinfo
  {author} {\bibfnamefont {P.}~\bibnamefont {S{\'a}nchez}},\ and\ \bibinfo
  {author} {\bibfnamefont {E.}~\bibnamefont {L{\'o}pez}},\ }\bibfield  {title}
  {\bibinfo {title} {Magnetic properties of permalloy multilayers with
  alternating perpendicular anisotropies},\ }\href
  {https://doi.org/10.1140/epjb/e2008-00163-4} {\bibfield  {journal} {\bibinfo
  {journal} {The European Physical Journal B}\ }\textbf {\bibinfo {volume}
  {62}},\ \bibinfo {pages} {267} (\bibinfo {year} {2008})}\BibitemShut
  {NoStop}%
\bibitem [{sup()}]{supplemental}%
  \BibitemOpen
  \href@noop {} {}\bibinfo {note} {See Supplemental Material at [URL will be
  inserted by publisher] for all XMCD-PEEM images with their graphical
  representation.}\BibitemShut {Stop}%
\bibitem [{\citenamefont {Erickson}\ \emph {et~al.}(1993)\citenamefont
  {Erickson}, \citenamefont {Prost},\ and\ \citenamefont
  {Timins}}]{Erickson1993}%
  \BibitemOpen
  \bibfield  {author} {\bibinfo {author} {\bibfnamefont {S.~J.}\ \bibnamefont
  {Erickson}}, \bibinfo {author} {\bibfnamefont {R.~W.}\ \bibnamefont
  {Prost}},\ and\ \bibinfo {author} {\bibfnamefont {M.~E.}\ \bibnamefont
  {Timins}},\ }\bibfield  {title} {\bibinfo {title} {The ``magic angle''
  effect: Background physics and clinical relevance.},\ }\href
  {https://doi.org/10.1148/radiology.188.1.7685531} {\bibfield  {journal}
  {\bibinfo  {journal} {Radiology}\ }\textbf {\bibinfo {volume} {188}},\
  \bibinfo {pages} {23} (\bibinfo {year} {1993})}\BibitemShut {NoStop}%
\bibitem [{\citenamefont {{de Sousa}}\ \emph {et~al.}(2004)\citenamefont {{de
  Sousa}}, \citenamefont {Delgado},\ and\ \citenamefont
  {Das~Sarma}}]{deSousa2004}%
  \BibitemOpen
  \bibfield  {author} {\bibinfo {author} {\bibfnamefont {R.}~\bibnamefont {{de
  Sousa}}}, \bibinfo {author} {\bibfnamefont {J.~D.}\ \bibnamefont {Delgado}},\
  and\ \bibinfo {author} {\bibfnamefont {S.}~\bibnamefont {Das~Sarma}},\
  }\bibfield  {title} {\bibinfo {title} {Silicon quantum computation based on
  magnetic dipolar coupling},\ }\href
  {https://doi.org/10.1103/PhysRevA.70.052304} {\bibfield  {journal} {\bibinfo
  {journal} {Physical Review A}\ }\textbf {\bibinfo {volume} {70}},\ \bibinfo
  {pages} {052304} (\bibinfo {year} {2004})}\BibitemShut {NoStop}%
\bibitem [{\citenamefont {Jensen}\ \emph {et~al.}(2022)\citenamefont {Jensen},
  \citenamefont {Str{\o}mberg}, \citenamefont {Lykkeb{\o}}, \citenamefont
  {Penty}, \citenamefont {Leliaert}, \citenamefont {Sj{\"a}lander},
  \citenamefont {Folven},\ and\ \citenamefont {Tufte}}]{Flatspin2022}%
  \BibitemOpen
  \bibfield  {author} {\bibinfo {author} {\bibfnamefont {J.~H.}\ \bibnamefont
  {Jensen}}, \bibinfo {author} {\bibfnamefont {A.}~\bibnamefont
  {Str{\o}mberg}}, \bibinfo {author} {\bibfnamefont {O.~R.}\ \bibnamefont
  {Lykkeb{\o}}}, \bibinfo {author} {\bibfnamefont {A.}~\bibnamefont {Penty}},
  \bibinfo {author} {\bibfnamefont {J.}~\bibnamefont {Leliaert}}, \bibinfo
  {author} {\bibfnamefont {M.}~\bibnamefont {Sj{\"a}lander}}, \bibinfo {author}
  {\bibfnamefont {E.}~\bibnamefont {Folven}},\ and\ \bibinfo {author}
  {\bibfnamefont {G.}~\bibnamefont {Tufte}},\ }\bibfield  {title} {\bibinfo
  {title} {Flatspin: A large-scale artificial spin ice simulator},\ }\href
  {https://doi.org/10.1103/PhysRevB.106.064408} {\bibfield  {journal} {\bibinfo
   {journal} {Physical Review B}\ }\textbf {\bibinfo {volume} {106}},\ \bibinfo
  {pages} {064408} (\bibinfo {year} {2022})}\BibitemShut {NoStop}%
\bibitem [{\citenamefont {M{\"o}ller}\ and\ \citenamefont
  {Moessner}(2009)}]{Moller2009}%
  \BibitemOpen
  \bibfield  {author} {\bibinfo {author} {\bibfnamefont {G.}~\bibnamefont
  {M{\"o}ller}}\ and\ \bibinfo {author} {\bibfnamefont {R.}~\bibnamefont
  {Moessner}},\ }\bibfield  {title} {\bibinfo {title} {Magnetic multipole
  analysis of kagome and artificial spin-ice dipolar arrays},\ }\href
  {https://doi.org/10.1103/PhysRevB.80.140409} {\bibfield  {journal} {\bibinfo
  {journal} {Physical Review B}\ }\textbf {\bibinfo {volume} {80}},\ \bibinfo
  {pages} {140409} (\bibinfo {year} {2009})}\BibitemShut {NoStop}%
\bibitem [{\citenamefont {M{\"o}ller}\ and\ \citenamefont
  {Moessner}(2006)}]{Moller2006}%
  \BibitemOpen
  \bibfield  {author} {\bibinfo {author} {\bibfnamefont {G.}~\bibnamefont
  {M{\"o}ller}}\ and\ \bibinfo {author} {\bibfnamefont {R.}~\bibnamefont
  {Moessner}},\ }\bibfield  {title} {\bibinfo {title} {Artificial square ice
  and related dipolar nanoarrays},\ }\href
  {https://doi.org/10.1103/PhysRevLett.96.237202} {\bibfield  {journal}
  {\bibinfo  {journal} {Physical Review Letters}\ }\textbf {\bibinfo {volume}
  {96}},\ \bibinfo {pages} {237202} (\bibinfo {year} {2006})}\BibitemShut
  {NoStop}%
\end{thebibliography}%

\end{document}


\title{Supplemental Material: Coupling-dependent Antiferromagnetic-Ferromagnetic Ordering in Pinwheel Artificial Spin Ice}

\author{Anders Str\o mberg}
 \email{anders.stromberg@ntnu.no}
 \affiliation{Department of Electronic Systems, Norwegian University of Science and Technology, Trondheim, Norway.}

 \author{Einar Digernes}
  \affiliation{Department of Electronic Systems, Norwegian University of Science and Technology, Trondheim, Norway.}

\author{Rajesh Vilas Chopdekar}
 \affiliation{Advanced Light Source, Lawrence Berkeley National Laboratory, Berkeley, CA, USA}

\author{Jostein Grepstad}%
 \affiliation{Department of Electronic Systems, Norwegian University of Science and Technology, Trondheim, Norway.}%
\author{Erik Folven}%
 \affiliation{Department of Electronic Systems, Norwegian University of Science and Technology, Trondheim, Norway.}%

\maketitle

\setcounter{figure}{0}
\renewcommand{\thefigure}{S\arabic{figure}}

\begin{figure*}[h]
    \centering
    \includegraphics[width=\textwidth]{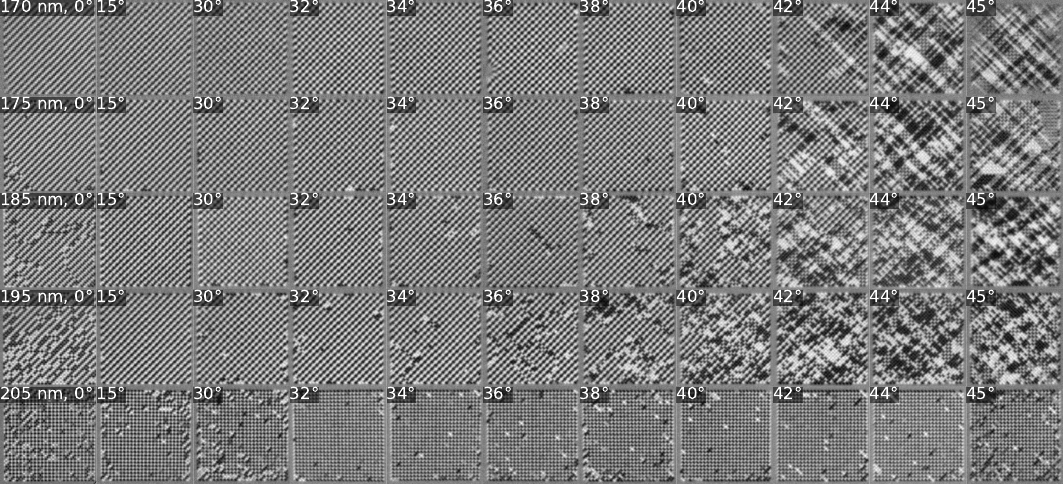}
    \caption{XMCD-PEEM images of the ASI systems in their relaxed state.
    All separations $d$ and rotation angles $\theta$ included in this paper are displayed.}
    \label{fig:cartoon-combined}
\end{figure*}

\begin{figure*}[h]
    \centering
    \includegraphics[width=\textwidth]{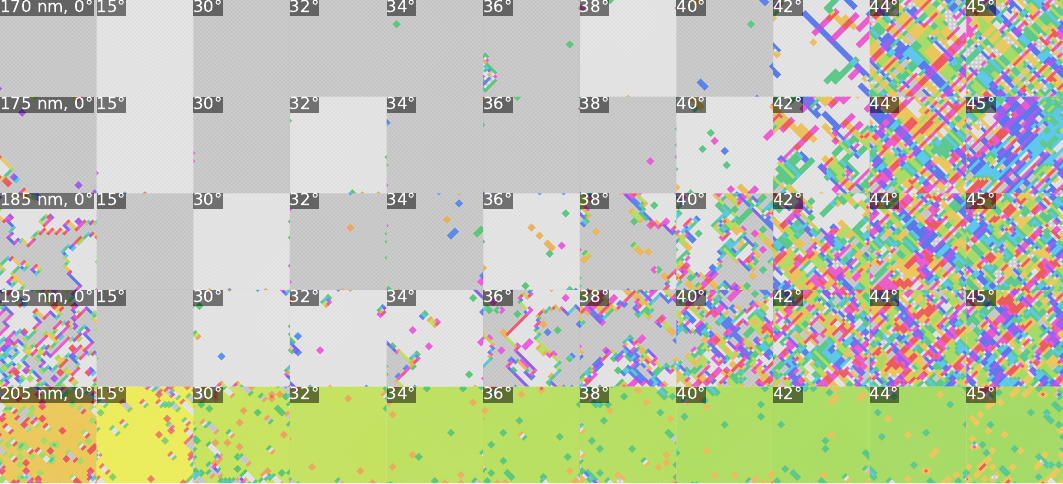}
    \caption{Analyzed magnetization maps of the ASI systems in their relaxed state. 
    All separations $d$ and rotation angles $\theta$ included in this paper are displayed.
    Gray domains indicate antiferromagnetically ordered regions, where each shade of gray indicates one of two possible antiferromagnetic phases. Colored regions are ferromagnetically ordered with a net magnetization as in the main paper.}
    \label{fig:cartoon-decoded}
\end{figure*}

\begin{figure*}[h]
    \centering
    \includegraphics[width=\textwidth]{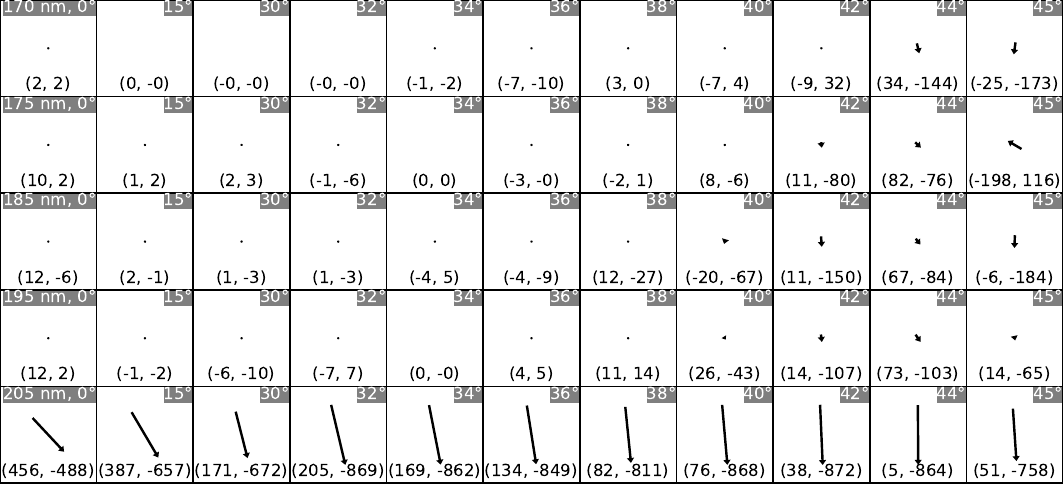}
    \caption{Total magnetization of the ASI systems in their relaxed state. 
    All separations $d$ and rotation angles $\theta$ included in this paper are displayed.
    The arrows indicate the size and direction of the total magnetization (where the arrows are too small, they appear as dots). 
    The numbers in parenthesis indicate the $x$ and $y$ components of the magnetization, in units of single spins, e.g., the numbers (8, -6) indicate that there is an excess of 8 magnets pointing in the positive $x$ direction and an excess of 6 magnets pointing in the negative $y$ direction.
   }
    \label{fig:cartoon-normfields}
\end{figure*}

\begin{figure*}[h]
    \centering
    \includegraphics[width=\textwidth]{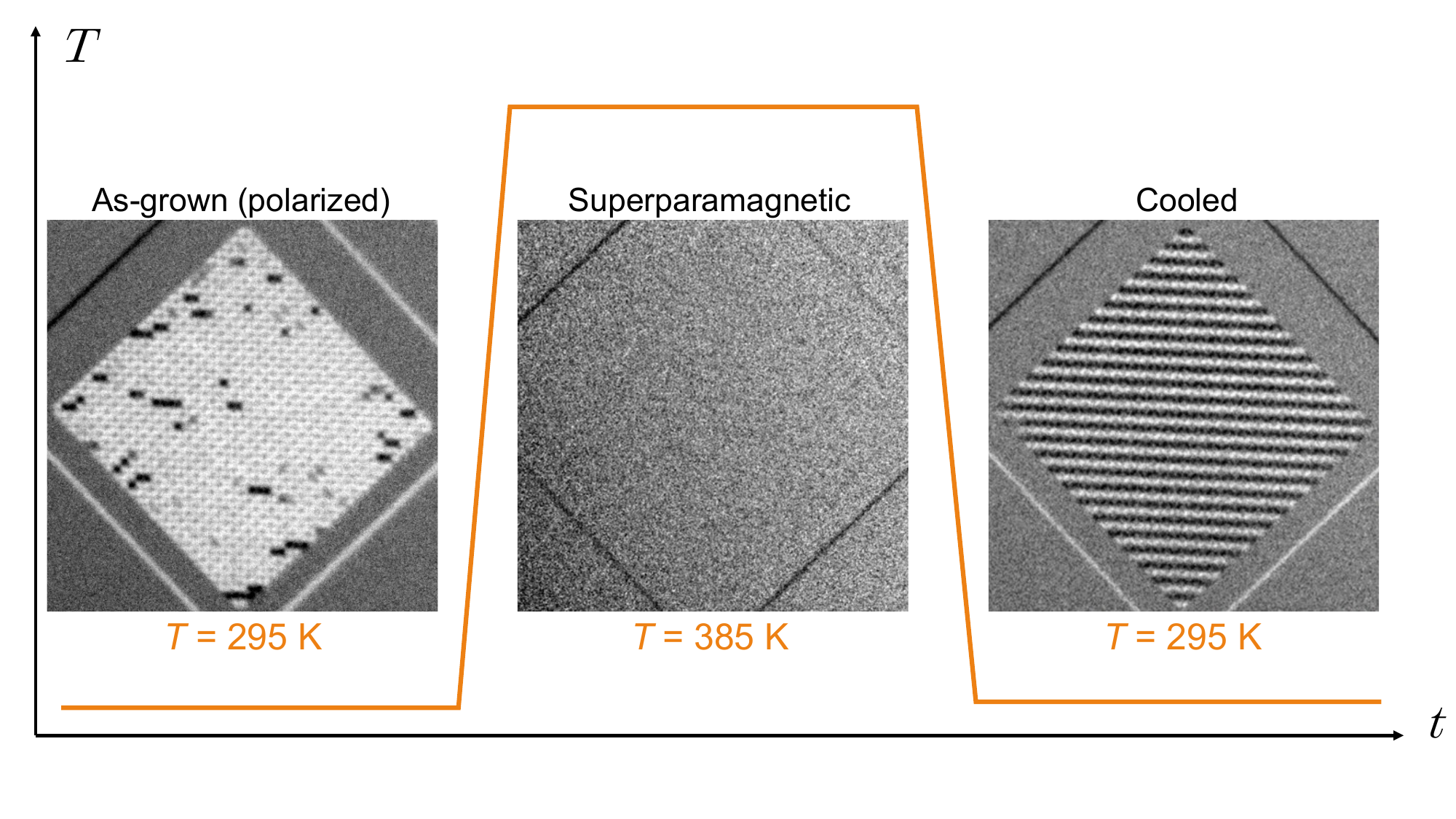}
    \caption{
    Illustration with XMCD-PEEM images of the magnetic state of a square ASI at different temperatures.
    White contrast indicates magnetization to the right, and black contrast to the left.
    The as-grown/polarized state is clearly ferromagnetic at room temperature.
    At $T = \SI{385}{\kelvin}$ the system is superparamagnetic, with no net magnetization over the imaging timeframe (although the constituent material is still ferromagnetic, as can be seen by the black frame lines).
    After the heat is switched on and the sample is returned to room temperature, it is clear that the system has ordered into a completely antiferromagnetic ground state.
    The orange temperature curve is just an illustration of the imaged temperatures $T$ with respect to the time sequence $t$, but not an actual recorded temperature curve over time. 
   }
    \label{fig:cartoon-normfields}
\end{figure*}